\pdfoutput=1
\RequirePackage{fix-cm}
\documentclass[twocolumn]{svjour3}          % twocolumn
\smartqed  % flush right qed marks, e.g. at end of proof
\usepackage{graphicx}
\usepackage{amsmath}
\usepackage[utf8]{inputenc} % allow utf-8 input
\usepackage[T2A,T1]{fontenc}  % use 8-bit T1 fonts, T2A Cyrillic fonts
\usepackage{lmodern}
\usepackage{url}            % simple URL typesetting
\usepackage{graphicx}       % more modern
\usepackage{xcolor}
\usepackage{color}
\usepackage{CJKutf8}
\usepackage[russian,english]{babel}
\newcommand{\ja}[1]{\begin{CJK}{UTF8}{min}#1\end{CJK}}

\usepackage[labelfont=bf,labelsep=quad]{caption}
\usepackage{booktabs}
\usepackage{threeparttable}
\usepackage{colortbl}
\usepackage{wrapfig}  % for floating
\usepackage{multirow}  % colspan/rowspan

\usepackage{amssymb}  % for \checkmark

\usepackage[title]{appendix}

\usepackage{enumitem}

\usepackage[unicode=true,hidelinks]{hyperref}

\usepackage{listings}
\lstdefinestyle{listingstyle}{basicstyle=\ttfamily\scriptsize}
\journalname{International Journal on Digital Libraries}
\begin{document}

\title{Cross-Lingual Citations in English Papers}
\subtitle{A Large-Scale Analysis of Prevalence, Usage, and Impact}

%\titlerunning{Short form of title}        % if too long for running head

\author{Tarek Saier%\orcidID{0000-0001-5028-0109}
        \and
        Michael F{\"a}rber%\orcidID{0000-0001-5458-8645}
        \and
        Tornike Tsereteli
}

%\authorrunning{Short form of author list} % if too long for running head

\institute{Karlsruhe Institute of Technology (KIT),
              Kaiserstr. 89,
              76133 Karlsruhe, Germany,
              \email{tarek.saier@kit.edu, michael.faerber@kit.edu}
\and University of Stuttgart,
              Pfaffenwaldring 5b, 70569 Stuttgart, Germany,
              \email{tornike.tsereteli@ims.uni-stuttgart.de}
}

\date{Received: date / Accepted: date}
% The correct dates will be entered by the editor

\maketitle

\begin{abstract}
Citation information in scholarly data is an important source of insight into the reception of publications and the scholarly discourse. Outcomes of citation analyses and the applicability of citation based machine learning approaches heavily depend on the completeness of such data. One particular shortcoming of scholarly data nowadays is that non-English publications are often not included in data sets, or that language metadata is not available.
Because of this, citations between publications of differing languages (cross-lingual citations) have only been studied to a very limited degree. In this paper, we present an analysis of cross-lingual citations based on over one million English papers, spanning three scientific disciplines and a time span of three decades.
Our investigation covers differences between cited languages and disciplines, trends over time, and the usage characteristics as well as impact of cross-lingual citations. Among our findings are an increasing rate of citations to publications written in Chinese, citations being primarily to local non-English languages, and consistency in citation intent between cross- and monolingual citations.
To facilitate further research, we make our collected data and source code publicly available.
% \PACS{PACS code1 \and PACS code2 \and more}
% \subclass{MSC code1 \and MSC code2 \and more}
\end{abstract}

\begin{figure}[tb]
\centering
\includegraphics[width=\linewidth]{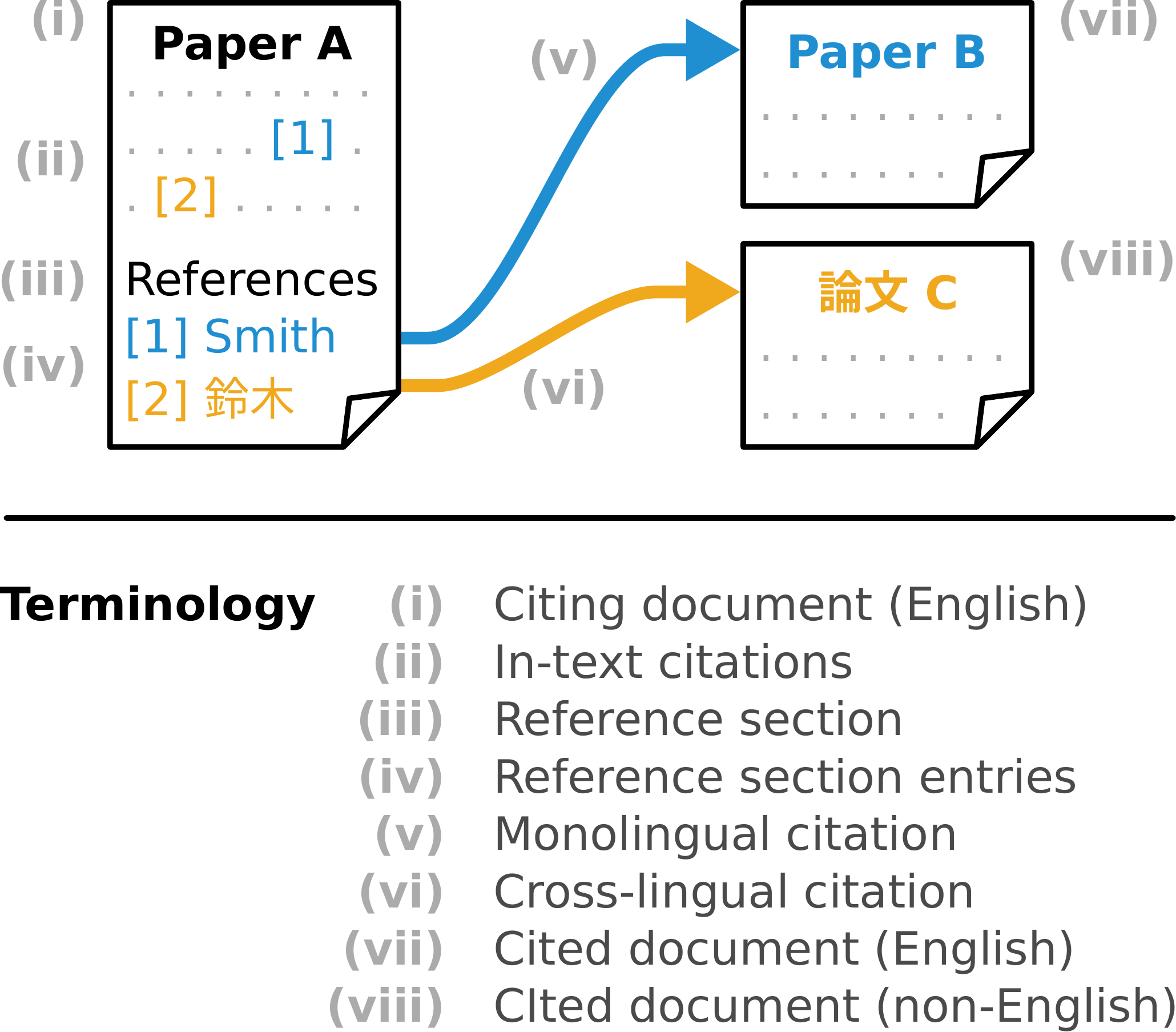}
\caption{Schematic explanation of terminology.} \label{fig:terminology}
\end{figure}

\section{Introduction}
Citations are an essential tool for scientific practice. By allowing authors to refer to existing publications, citations make it possible to position one's work within the context of others', critique, compare, and point readers to supplementary reading material. In other words, citations enable scientific discourse. Because of this, citations are a valuable indicator for the academic community's reception of and interaction with published works. Their analysis is used, for example, to quantify research output~\cite{Hirsch2005}, qualify references~\cite{Abujbara2013}, and detect trends~\cite{Chen2006}. Furthermore, citations can be utilized to aid researchers through, for example, summarization~\cite{Elkiss2008} or recommendation~\cite{Ma2020,Faerber202x} of papers, and through applications driven by document embeddings in general~\cite{Cohan2020}.

As such analyses and applications require data to be based on, the availability of citation data or lack thereof is decisive with regard to the areas, in which respective insights can be gained and approaches developed. Here, the literature points 
in two major directions of lacking coverage---namely the humanities~\cite{Colavizza2019,Kellsey2004} and non-English publications~\cite{Vera-Baceta2019,Liu2019,Moed2018,Moskaleva2019}.
Because most large scholarly data sets are either artificially limited to few languages (e.g., English only) or do not provide language metadata,
a particular practice not well researched so far is cross-lingual citation. That is, references where the citing and cited documents are written in different languages (see \textit{(vi)} in Figure~\ref{fig:terminology}).
Cross-lingual citations are, however, important bridges between otherwise insufficiently connected ``language silos''~\cite{Shu2019,Moskaleva2019}.

Because English is currently the de facto academic lingua franca~\cite{Montgomery2013}, citations from non-English languages to English are significantly more prevalent than the other way around. This dichotomy is reflected in existing literature, where usually either citations from English~\cite{Kellsey2004,Lillis2010}, or to English~\cite{Tang2014,Jiang2018,Jiang2018b,Schrader2019} are analyzed. As both directions involve a non-English document on one side of the citation, the analysis of either is challenging with today's Anglocentric state of citation data.

Setting our focus to cross-lingual citations \emph{from English}, we perform a large-scale analysis on over one million documents.
In line with existing literature we determine the prevalence of cross-lingual citations across multiple dimensions. Additionally, we investigate the citation's usage as well as impact. In particular, the following research questions are addressed.

\begin{enumerate}[align=left]
\item[RQ1)] How prevalent are English to non-English references? We consider prevalence in general, in different disciplines, across time, and within publications that use them.
\item[RQ2)] In what circumstances are cross-lingual citations in English papers used? Here we consider self-citation, geographic origin, as well as citation function and sentiment.
\item[RQ3)] What is the impact of cross-lingual citations in English documents? We consider the aspects of acceptance, data mining challenges, as well as impact on the success of a publication.
\end{enumerate}

\noindent Through our analysis, we make the following contributions.

\begin{enumerate}
\item We conduct an analysis of cross-lingual citations in English papers that is considerably more extensive than existing literature in terms of corpus size as well as covered languages, time, and disciplines. This not only makes the results more representative of the areas covered, but also enables the use of our collected data for machine learning based applications such as cross-lingual citation recommendation.
\item We propose an easy and reliable method for identifying cross-lingual citations from English papers to publications in non-Latin script languages (e.g., Russian and Chinese).
\item We highlight key challenges for handling cross-lingual citations that can inform future developments in scholarly data mining.
\item To facilitate further research, we make our collected data, source code, and full results publicly available.\footnote{See \url{https://github.com/IllDepence/cross-lingual-citations-from-en}.}
\end{enumerate}

\noindent The remainder of the paper is structured as follows. After briefly addressing our use of terminology down below, we give an overview of related work in Section~\ref{sec:relwork}. In Section~\ref{sec:datacollection} we discuss the identification of cross-lingual citations, data sources considered, and our data collection process. Subsequent analyses with regard to our research questions are then covered in Section~\ref{sec:results}. We end with a discussion of our findings and concluding remarks in Section~\ref{sec:conclusion}.

\subsection*{Terminology}
Because \textit{citation}, \textit{reference} and related terms are not used consistently in literature, we briefly address their use in this paper. As shown in Figure~\ref{fig:terminology}, a cit\emph{ing} document creates a bibliographical link to a cit\emph{ed} document. We use the terms \textit{citation} and \textit{reference} interchangeably for this type of link (e.g., ``\textit{(vi)} in Figure~\ref{fig:terminology} marks a cross-lingual reference,'' or ``Paper$^a$ makes two citations''). The textual manifestation of a bibliographic reference, often found at the end of a paper (e.g., ``[1] Smith'' in Figure~\ref{fig:terminology}), is referred to as \textit{reference section entry}, or sometimes \textit{reference} for short. We call the combined set of these entries \textit{reference section}. Lastly, parts within the text of a paper, which contain a marker connected to one of the reference section entries are called \textit{in-text citations}.

\section{Related Work}\label{sec:relwork}

\begin{table*}
\caption{Comparison of corpora}
 \label{tab:relworkcomp}
  \centering
  \begin{small}
 \begin{threeparttable}
 \begin{tabular}{lcrrrr}
 \toprule
   Work & Type\tnote{a} & \#Documents & \hphantom{m}\#References & \hphantom{m}\#Years & \hphantom{nn}\#Disciplines \\
   \midrule
   Kellsey and Knievel~\cite{Kellsey2004} & en$\rightarrow$* & 468 & 16k & 5\tnote{b} & 4 \\
   Lillis et al.~\cite{Lillis2010} & en$\rightarrow$* & 240 & 10k & 7 & 1 \\
   Schrader~\cite{Schrader2019} & *$\rightarrow$en & 403 & 5k & 2 & 1 \\
   Tang et al.~\cite{Tang2014} & zh$\rightarrow$en & 2k & 17k & 10 & 1 \\
   Jiang et al.~\cite{Jiang2018,Jiang2018b} & zh$\rightarrow$\{en,zh\} & 14k & 38k & n/a & 1 \\
   Kirchik et al.~\cite{Kirchik2012} & \{en,ru\}$\rightarrow$ru & 497k & n/a & 17 & (unrestricted) \\
   Ours & en$\rightarrow$* &  1.1M & 39M & 27 & 3 \\
   \bottomrule
 \end{tabular}
 \begin{tablenotes}
    \item[a] type$=$focus reference type (en$=$English, ru$=$Russian, zh$=$Chinese, *$=$any)
    \item[b] over a span of 40 years
 \end{tablenotes}
\end{threeparttable}
  \end{small}
\end{table*}

Existing literature on cross-lingual citations in academic publications covers analyses as well as approaches to prediction tasks. These are, however, only based on small corpora or restricted to specific language pairs. As shown in Table~\ref{tab:relworkcomp}, our work is based on a considerably larger corpus which is also more comprehensive in terms of the time span and disciplines that are covered.

In the following, we describe the works in Table~\ref{tab:relworkcomp} in more detail, reporting on the key corpus characteristics and findings. This is complemented by a short overview of existing literature on various types of cross-lingual interconnections in media other than academic publications.

\subsection{Cross-Lingual Citations in Academic Publications}

Literature concerning cross-lingual citations in academic publications can be found in the form of analyses and applications. In \cite{Kellsey2004} Kellsey and Knievel conduct an analysis of 468 articles containing 16,138 citations. The analysis spans 4 English language journals in the humanities (disciplines: history, classics, linguistics, and philosophy) over 5 particular years (1962, 1972, 1982, 1992, and 2002). They count cross-lingual citations to English, German, French, Italian, Spanish, Portuguese, and Latin, while further languages are grouped into a category ``other.''
The authors find that 21.3\% of the citations in their corpus are cross-lingual, but note strong differences between the covered disciplines. Over time, they observe a steady total, but declining relative number of cross-lingual citations per article. The authors furthermore find, that the ratio of publications that contain at least one cross-lingual citation is increasing.

Lillis et al.~\cite{Lillis2010} investigate if the global status of English is impacting the ``citability'' of non-English works in English publications. They base their analysis on 240 articles from 2000 to 2007 in psychology journals, and furthermore use the Social Sciences Citation Index and ethnographic records. Their corpus contains 10,688 references, of which 8.5\% are cross-lingual. Analyzing the prevalence of references in various contexts, they find that authors are more likely to cite a ``local language'' in English-medium national journals than in international journals. Further conducting analyses of e.g. in-text citation surface forms, they come to the conclusion that there are strong indicators for a pressure to cite English rather than non-English publications.

Similar observations are made by Kirchik et al.~\cite{Kirchik2012} concerning citations to Russian. Analyzing 498,221 papers in Thomson Reuters' Web of Science between 1993 and 2010, they find that Russian scholars are more than twice as likely to cite Russian publications when publishing in Russian language journals (21\% of citations) than when they publish in English (10\% of citations).

In \cite{Schrader2019} Schrader analyzes citations from non-English documents to English articles in open access and ``traditional'' journals. The corpus used comprises 403 cited articles published between 2011 and 2012 in the discipline of library and information science. The articles were cited 5,183 times (13.8\% by non-English documents). In their analysis the author observes that being open access makes no statistically significant difference for the ratio of incoming cross-lingual citations of an article, or the language composition of citations a journal receives.

Apart from analyses, there are also approaches to prediction tasks based on cross-lingual citations~\cite{Tang2014,Jiang2018,Jiang2018b,Ma2020}. Tang et al.~\cite{Tang2014} propose a bilingual context-citation embedding algorithm for the task of predicting suitable citations to English publications in Chinese sentences. To train and evaluate their approach, they use 2,061 articles from 2002 to 2012 in the Chinese Journal of Computers, which contain citations to 17,693 English publications. Comparing to several baseline methods, they observe the best performance for their novel system. Similarly, in \cite{Jiang2018} and \cite{Jiang2018b} Jiang et al. propose two novel document embedding methods jointly learned on publication content and citation relations. The corpus used in both cases consists of 14,631 Chinese computer science papers from the Wanfang digital library. The papers contain 11,252 references to Chinese publications and 27,101 references to English publications. For the task of predicting a list of suitable English language references for a Chinese query document, both approaches are reported to outperform a range of baseline methods.

\subsection{Cross-Lingual Interconnections in Other Types of Media}

Apart from academic publications, cross-lingual connections are also described in other types of media. Hale~\cite{Hale2012} analyzes cross-lingual hyperlinks between online blogs centered around a news event in 2010. In a corpus of 113,117 blog pages in English, Spanish, and Japanese, 12,527 hyperlinks (5.6\% of them cross-lingual) are identified. Analysis finds that less than 2\% of links in English blogs are cross-lingual, while the number in Spanish and Japanese blogs is slightly above 10\%. Hyperlinks between Spanish and Japanese are almost non-existent (7 in total). Further investigating the development of links over time, the author observes a gradual decrease in language group insularity driven by individual translations of blog content---a phenomenon described as ``bridgeblogging'' by Zuckerman~\cite{Zuckerman2008}.
Similar structural features are reported by Eleta et al.~\cite{Eleta2012} and Hale~\cite{Hale2014a} for Twitter, where multilingual users are bridging language communities.

Focusing on types of information diffusion that are not textually manifested through connections such as bibliographic references and hyperlinks, there also is literature on cross-lingual phenomena on collaborative online platforms, such as the study of cross-lingual information diffusion on Wikipedia~\cite{Kim2016,Samoilenko2016}.

Lastly, as with academic publications, there furthermore exists literature on link prediction tasks. In \cite{Jin2017} Jin et al. analyze cross-lingual information cascades and develop a machine learning approach based on language and content features to predict the size and language distribution of such cascades.

\section{Data Collection}\label{sec:datacollection}

In this section, we first discuss how to identify cross-lingual citations. Subsequently, we outline the steps of data source selection and corpus construction. Lastly, we describe the key characteristics of our corpus.

\subsection{Identification of Cross-Lingual Citations}
\label{sec:ident}
Identifying cross-lingual citations requires information about the language of the citing and cited document. However, this is often missing in scholarly data sets.\footnote{Details are provided in Section~\ref{sec:dataselect}.} Identifying the involved documents' language when it is not given in metadata, however, is challenging, because (a) the full text, especially of the cited documents, is not always available, (b) abstracts are not reliable because non-English publications often provide an additional English abstract, and (c) language identification on short strings (e.g., titles in references) does not achieve sufficient results with existing techniques~\cite{Jauhiainen2019}.

To nevertheless be able to conduct an analysis of cross-lingual citations on a large scale, we utilize the common practice of authors appending an explicit marker in the form of \textit{``(in $<$Language$>$)''} to such references. This shifts the requirements from language metadata or language identification to the existence of reference section entries in the data. This is because the language of the cited document is given by the \textit{``$<$Language$>$''} part of the marker, and the language the marker itself is written in (i.e., English) provides the citing document's language. For example, the reference section entry \textit{``M. Saitou, `Hydrodynamics on non-commutative space' (in Japanese), [...]''}\footnote{Found in \texttt{arXiv:1612.01831}.} by itself contains enough information to determine that the cited document is written in Japanese and the citing document is written in English.

The question then remains, how common the practice of using such explicit markers is---that is, to cite, for example, \textit{``A Modern Model Description of Magnetism (in Russian)''} instead of \textit{``\foreignlanguage{russian}{Современное модельное описание магнетизма}''}.\footnote{Referring to \texttt{arXiv:1103.5123}.} To answer this question, we perform a preliminary analysis on the data set unarXive~\cite{Saier2020}, which comprises 39 million reference section entries. Specifically, we conduct a large automated analysis on all reference section entries in the data set and additionally perform a smaller, manual analysis on a stratified sample of 5,000 references.

In the large automated analysis, we first identify the cited document's title within references using the state-of-the-art~\cite{Tkaczyk2018} reference string parser module of GROBID~\cite{Lopez2009}, and then determine the title's language using the language identification tool Lingua,\footnote{See \url{https://github.com/pemistahl/lingua}.} which is specialized for very short text.
Manually inspecting our results, we note that non-Latin script languages (e.g., Chinese, Japanese, Russian) are detected reliably,\footnote{To be more precise, no language that uses a script different to the Latin alphabet appears to be falsely identified as English. We are, however, not able to judge whether languages using the same non-Latin script---such as languages written in Cyrillic---are distinguished correctly by Lingua.} but Latin script languages (e.g., German and French) are not. For instance, many English titles are falsely identified as German.

\begin{table}
\caption{References to non-Latin script languages in the automated analysis}
 \label{tab:automatedresults}
  \centering
  \begin{small}
 \begin{threeparttable}
 \begin{tabular}{lrr}
 \toprule
   Cited Language & \#marked & \#unmarked \\
   \midrule
   Russian & 23,922 & 303 (1.3\%)\\
   Chinese & 2,351 & 10 (0.4\%) \\
   Japanese & 1,843 & 5 (0.3\%) \\
   Ukrainian & 876 & 15 (1.7\%)\\
   Bulgarian & 67 & 0 (0.0\%)\\
   Greek & 60 & 1 (1.7\%) \\
   \bottomrule
 \end{tabular}
\end{threeparttable}
  \end{small}
\end{table}

For non-Latin script languages, which we is shown in Table~\ref{tab:automatedresults}, only a small fraction of cross-lingual citations is not explicitly marked. We observe ratios of unmarked cross-lingual citations relative to explicit markers consistently below 2\%.\footnote{Because our analysis is based on language identification of the titles of cited publications, we cannot detect when a non-English work is cited with a translated title \emph{and} no explicit language marker.}

\begin{table}
\caption{Results of manual labeling}
 \label{tab:manualresults}
  \centering
  \begin{small}
 \begin{threeparttable}
 \begin{tabular}{lrr}
 \toprule
   Cited Language & \#references & \#marked \\
   \midrule
   (n/a)\tnote{a} & 2,737 & 0 \\
   English & 2,188 & 0 \\
   French & 33 & 1 \\
   German & 27 & 0 \\
   Russian & 8 & 6\tnote{b}\\
   Italian & 5 & 1 \\
   Chinese & 1 & 1 \\
   Japanese & 1 & 1 \\
   \bottomrule
 \end{tabular}
 \begin{tablenotes}
    \item[a] These references did not contain the title of the cited document, which is common in physics papers.
    \item[b] The two remaining unmarked references contained the cited publication's title only transliterated into the Latin alphabet.
  \end{tablenotes}
\end{threeparttable}
  \end{small}
\end{table}

To get a reliable estimate for Latin script languages as well, we additionally perform a smaller, manual analysis. To this end, we label a stratified sample\footnote{The sample was stratified according to the referencing document's discipline and month of publication.} of 5,000 references from unarXive with the reference's language as well as whether an explicit language marker was used or not. The results of our evaluation are shown in Table~\ref{tab:manualresults}.
In accordance with our automated large analysis, we observe that non-Latin script languages are generally explicitly marked. For Latin script languages, however, explicit marking appears to be considerably less common. We additionally evaluate the automated language identification results for our manually annotated references and measure F1 scores of 0.48, 0.46, and 0.60 for French, German, and Italian respectively. Notably, less than half of the references with German titles are detected (44\% recall) and more than half of the references identified as German are false positives (48\% precision).

The results of above preliminary investigations have two consequences for the findings in our main analyses, which are based on explicit language markers. First, a direct comparison between our results on non-Latin and Latin script languages is only valid for \emph{explicitly marked} cross-lingual citations, as there is a notable amount of undetected cross-lingual citations for Latin script languages. Second, the number of undetected cross-lingual citations for non-Latin script languages such as Chinese, Japanese, and Russian, is negligible. Accordingly, concerning these languages, our results are valid for cross-lingual citations \emph{regardless of language markers}.

\subsection{Data Source Selection}\label{sec:dataselect}

\begin{table*}
\caption{Overview of data sets}
 \label{tab:datasets}
  \centering
  \begin{small}
 \begin{threeparttable}
 \begin{tabular}{lrlllc}
 \toprule
   Data set & \#Documents & Language metadata & Refs. resolved to & Reference sections & Used \\
   \midrule
   MAG\tnote{a}~~\cite{Sinha2015,Wang2019} & 230M  & (48\%\tnote{b}~) & MAG & - & \checkmark\\
   CORE\tnote{c} & 123M & 1.79\% & CORE & - & \\
   S2ORC~\cite{Lo2020} & 81M & - & S2ORC & 34\% (in GROBID parse) & \\
   PubMed Central OAS\tnote{d} & 2M & - & mixed & 100\% (in JATS XML) & \\
   unarXive~\cite{Saier2020} & 1M & - & MAG & 100\% (dedicated entity) & \checkmark\\
   \bottomrule
 \end{tabular}
 \begin{tablenotes}
    \item[a] Using version 2019-12-26
    \item[b] Language given for source URLs (not always matching paper language)
    \item[c] See \url{https://core.ac.uk/}. Using version 2018-03-01
    \item[d] See \url{https://www.ncbi.nlm.nih.gov/pmc/tools/openftlist/}
  \end{tablenotes}
\end{threeparttable}
  \end{small}
\end{table*}

As our data source we considered five large scholarly data sets commonly used for citation related tasks~\cite{Khan2017,Faerber202x}. Table~\ref{tab:datasets} gives an overview of their key properties. The Microsoft Academic Graph (MAG) and CORE are both very large data sets with some form of language metadata present. In the MAG the language is given not for documents themselves, but for URLs associated with papers. CORE contains a language label for 1.79\% of its documents. S2ORC, the PubMed Central Open Access Subset (PMC OAS), and unarXive do not offer language metadata, but all contain some form of reference sections (GROBID output, JATS~\cite{Huh2014} XML, and raw strings extracted from \LaTeX~source files respectively).

From these five, we decided to use unarXive and the MAG. This decision was motivated by two key reasons: (1)~metadata of cited documents, and (2)~evaluation of the acceptance of cross-lingual citations in English papers. As for (1), both S2ORC and the PMC OAS link references in their papers to document IDs within the data set itself (only partly in the PMC OAS, where also MEDLINE IDs and DOIs are found~\cite{Gipp2015}). This is problematic in our case, because S2ORC is restricted to English papers, and the PMC OAS is constrained to Latin script contents,\footnote{See \url{https://www.ncbi.nlm.nih.gov/pmc/about/faq/\#q16}.} which means metadata on non-English cited documents is non-existent (S2ORC) or very limited (PMC OAS). In unarXive, on the other hand, references are linked to the MAG, which contains metadata on publications regardless of language. Concerning reason (2), the fact that unarXive is built from papers on the preprint server arxiv.org, and the MAG contains metadata on paper's preprint \emph{and} published versions, allows us to analyze whether or not cross-lingual citations are affected by the peer review process.

With these two data sources selected, the extent of our analysis is over one million documents, across 3 disciplines (physics, mathematics, computer science), over a span of 27 years (1992--2019).

\subsection{Data Collection}\label{sec:datacollectionsub}
To identify references with \textit{``(in $<$Language$>$)''} markers, we iterate through the total of 39.7M reference section entries in unarXive and first filter for the regular expression \verb|\(\s*in\s+[a-zA-Z][a-z]+\s*\)|. This yields 51,380 matches with 207 unique tokens following \textit{``in''} within the parentheses. Within these 207 tokens we manually remove those referring to non-languages (e.g., ``press'' or ``preparation'') and correct misspellings (e.g., ``japanease'' or ``russain''), resulting in 44 unique language tokens. These are (presented in ISO 639-1 codes) be, bg, ca, cs, da, de, el, en, eo, es, et, fa, fi, fr, he, hi, hr, hu, hy, id, is, it, ja, ka, ko, la, lv, mk, mr, nl, no, pl, pt, ro, ru, sa, sk, sl, sr, sv, tr, uk, vi, and zh. These 44 languages cover 43 of the 78 languages, in which journals indexed in the Directory of Open Access Journals\footnote{See \url{https://doaj.org/}.} (DOAJ) are published as of July 2020. The one language found in our data, but with no journal in the DOAJ, is Marathi. In terms of journal count by language, above 44 languages cover 97.54\% of the DOAJ. In total, our data contains 33,290 reference section entries in 18,171 unique citing documents. We refer to this set of documents as the \emph{cross-lingual set}.

To analyze differences between papers containing cross-lingual citations in unarXive and a comparable random set, we also generate a second set of papers. To ensure comparability we go through each year of the cross-lingual set, note the number of documents per discipline and then randomly sample the same number of documents from all of unarXive within this year and discipline. This means the \emph{cross-lingual set} and the \emph{random set} have the same document distribution across years and disciplines. Table \ref{tab:dataused} gives an overview of the resulting data used.

\section{Results}\label{sec:results}

In the following we describe the results of our analyses with regard to the research questions laid out in the introduction. We begin with general numbers concerning the \emph{prevalence} of cross-lingual citations. These results are based on unarXive alone. This is followed by more in depth observations regarding cross-lingual citations' \emph{usage} (e.g., the underlying motivation or the citation's function) and \emph{impact} (e.g., acceptance by reviewers or challenges for data mining). These subsequent in depth analyses additionally utilize the MAG metadata.

\begin{table}
\caption{Overview of data used}
 \label{tab:dataused}
  \centering
  \begin{small}
 \begin{threeparttable}
 \begin{tabular}{lrrr}
 \toprule
   \multicolumn{2}{r}{Cross-lingual set} & Random set & unarXive \\
   \midrule
   \#Docs & 18,171 & 18,171 & 1,192,097 \\
   \#Docs (MAG) & 16,300 & 16,464 & 1,087,765 \\
   \#Refs & 635,154 & 536,672 & 39,694,083 \\
   \#Refs (MAG) & 290,421 & 242,090 & 15,954,664 \\
   \#Cross-lingual refs & 33,290 & 642 & 33,290 \\
   \bottomrule
 \end{tabular}
 \begin{tablenotes}
    \item *docs  =  documents,\\\hphantom{*}refs  =  reference section entries,\\\hphantom{*}(MAG) = with a MAG ID.
  \end{tablenotes}
\end{threeparttable}
  \end{small}
\end{table}

\subsection{Prevalence}

We find \textit{``(in $<$Language$>$)''} markers in 33,290 out of 39,694,083 reference section entries (0.08\%). These appear in 18,171 out of 1,192,097 documents (1.5\%)---in other words in every 66th document. Of these 18k documents, 17,223 cite one language other than English, 864 cite two, 76 three, 7 documents four, and a single document cites works in English and five further languages (Russian, French, Polish, Italian, and German). The five most common language pairs within a single document are Russian-Ukrainian (277 documents), German-Russian (166), French-Russian (135), French-German (68), and Chinese-Russian (59).

\begin{table}
\caption{Most prevalent languages}
 \label{tab:prevlangs}
  \centering
  \begin{small}
 \begin{threeparttable}
 \begin{tabular}{lrr}
 \toprule
   Language & \#References & \#Documents \\
   \midrule
   Russian & 23,922 & 12,304 \\
   Chinese & 2,351 & 1,582 \\
   Japanese & 1,843 & 1,397 \\
   German & 1,244 & 965 \\
   French & 931 & 719 \\
   \bottomrule
 \end{tabular}
\end{threeparttable}
  \end{small}
\end{table}

\begin{figure}[tb]
\includegraphics[width=\linewidth]{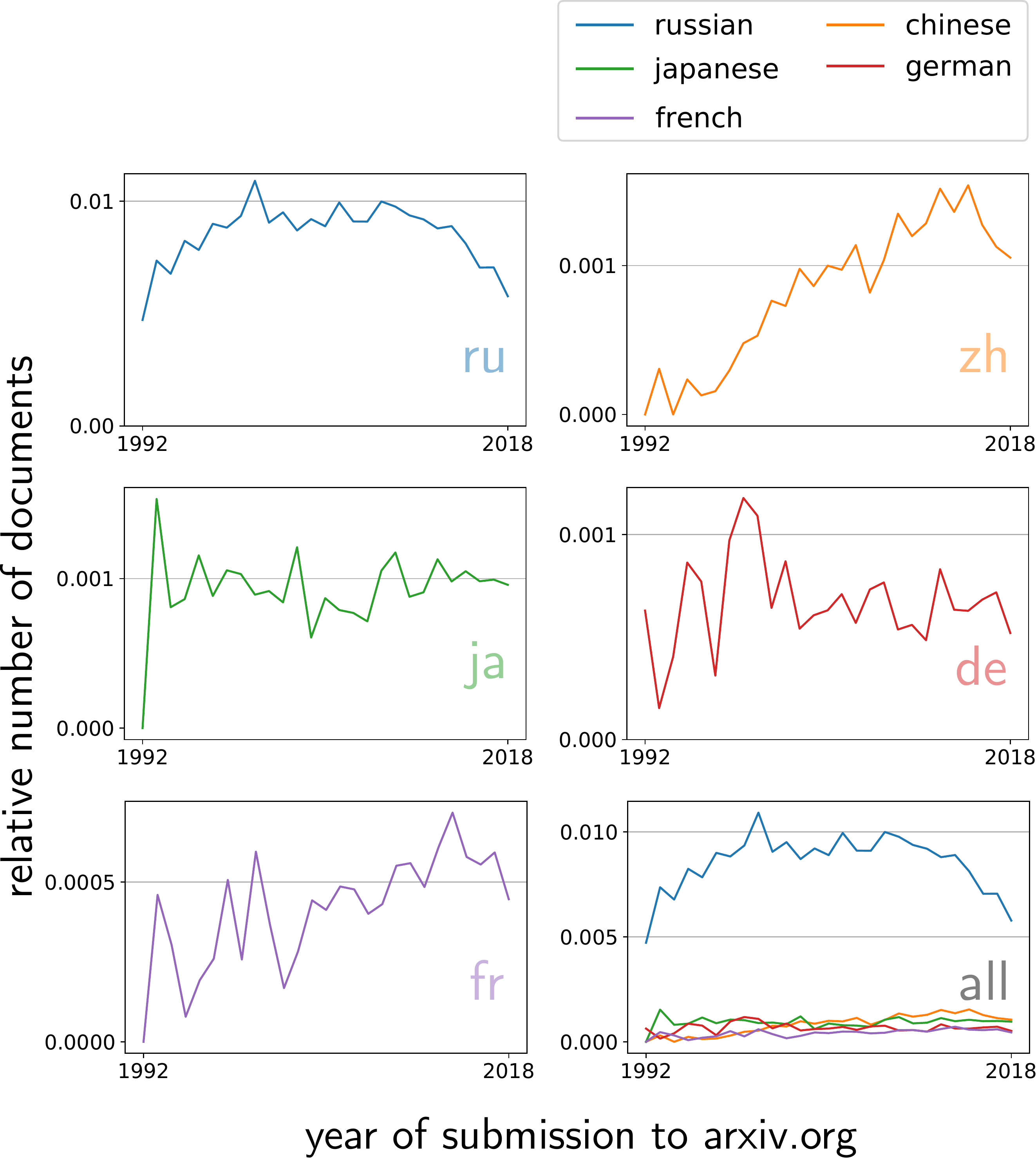}
\caption{Relative number of documents citing Russian, Chinese, Japanese, German, and French works. Showing all aforementioned in the bottom right.} \label{fig:previntime}
\end{figure}

\begin{figure}[tb]
\includegraphics[width=\linewidth]{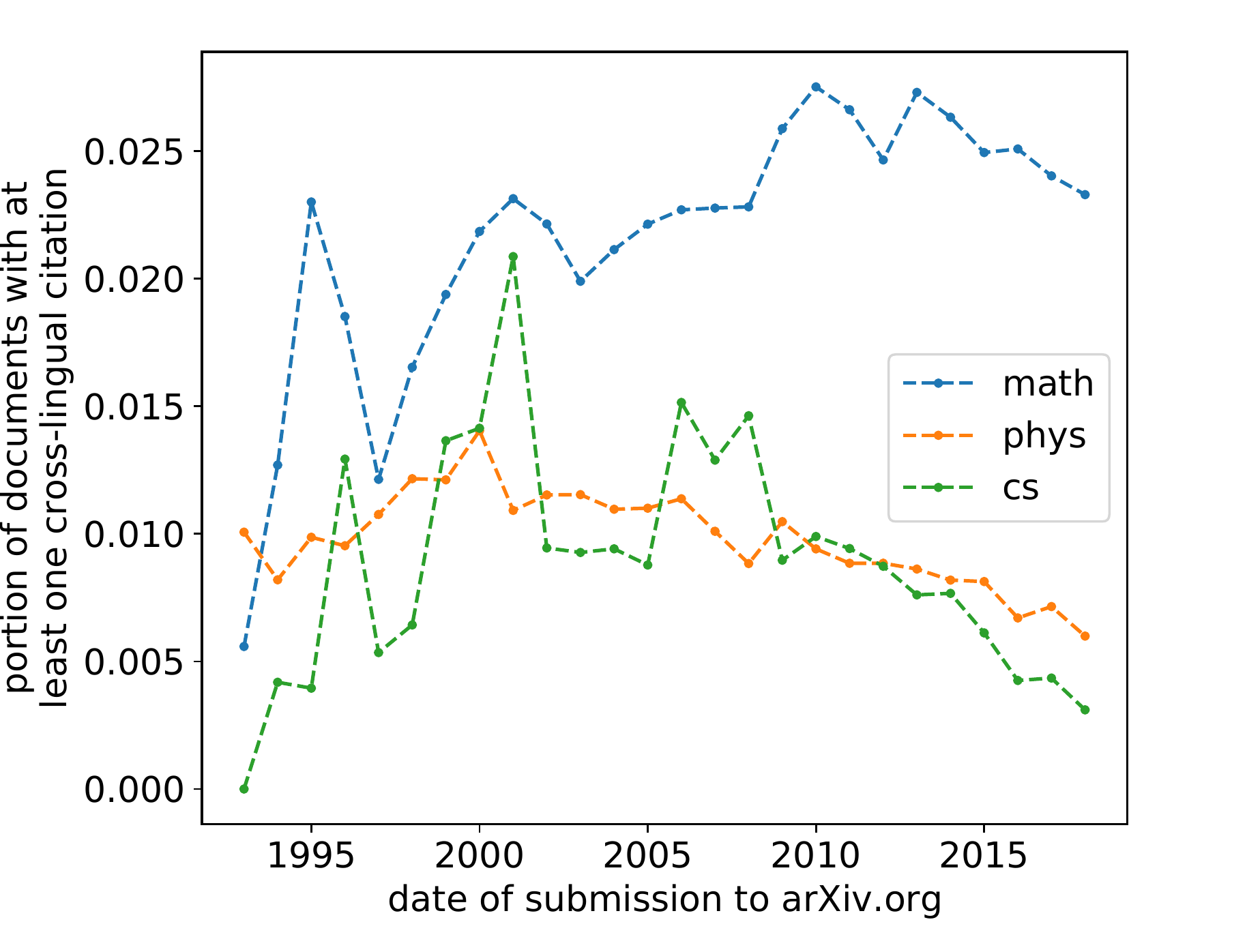}
\caption{Relative number of mathematics, physics, and computer science documents citing non-English works.} \label{fig:prevperdisc}
\end{figure}

Table~\ref{tab:prevlangs} shows the absolute number of reference section entries and unique citing documents for the five most prevalent languages, which combined make up over 90\% in terms of both references and documents. As we can see, Russian is by far the most common, making up about two thirds of the cross-lingual set. When breaking down these numbers by year or discipline, it is important to also factor in the distribution of documents along these dimensions in the whole data set. Doing so, we show in Figure~\ref{fig:previntime} the relative number of documents with cross-lingual citations over time for each of the aforementioned five languages. While the numbers in earlier years can be a bit unstable due to low numbers of total documents, we can observe a downwards trend of citations to Russian, an upwards trend of citations to Chinese, and a somewhat stable proportion in documents citing Japanese works. Looking at the numbers per discipline in Figure~\ref{fig:prevperdisc}, we can see that cross-lingual citations occur most often in mathematics papers, and are about half as common in physics and computer science.

\begin{figure}[tb]
\includegraphics[width=\linewidth]{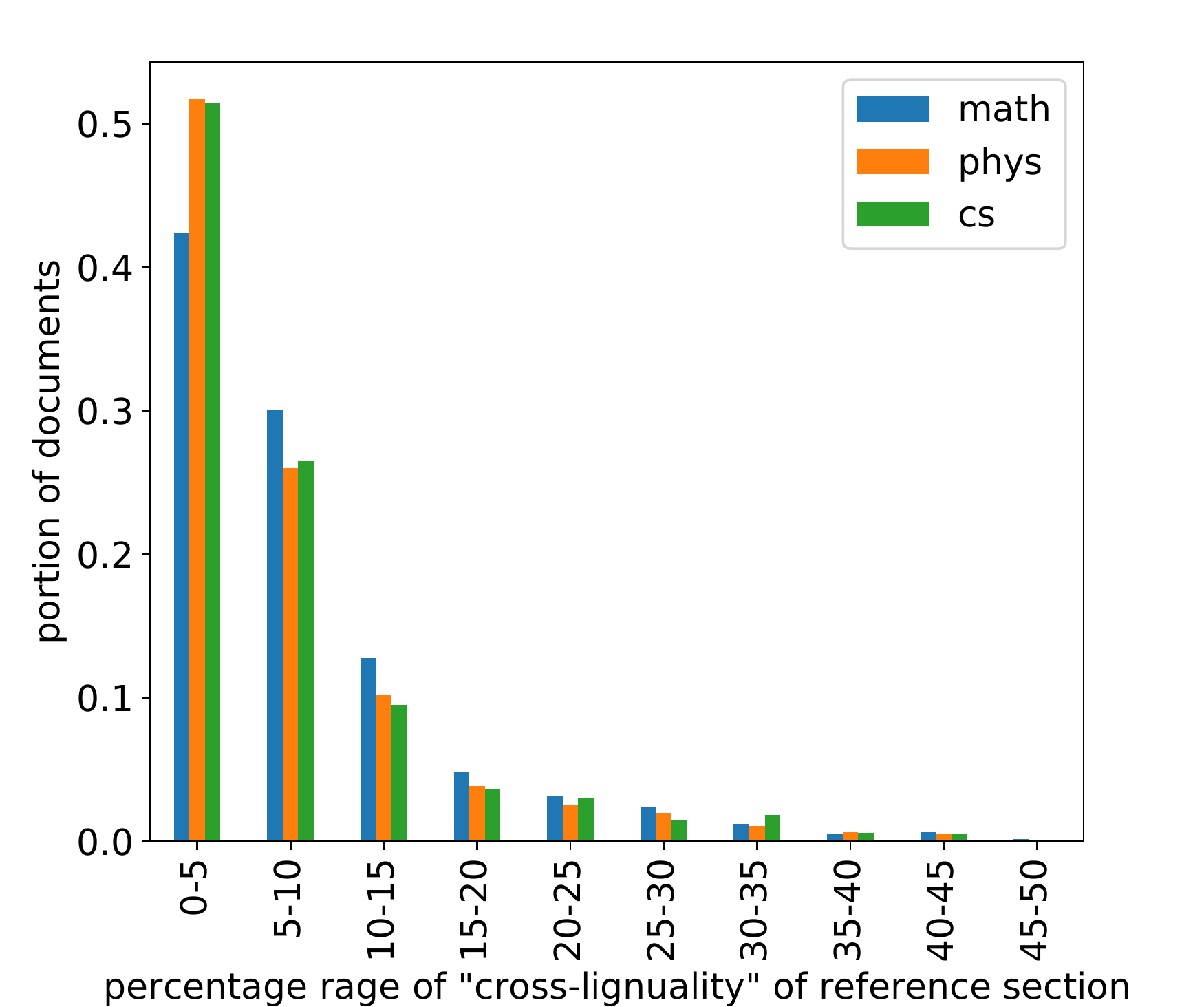}
\caption{``Cross-linguality'' of reference sections by discipline.} \label{fig:xlinglyhistbydisc}
\end{figure}

Lastly, within the reference section of a document that has at least one cross-lingual citation, the mean value of ``cross-linguality'' (i.e., what portion of the reference section is cross-lingual) is 0.083 with a standard deviation of 0.099. Breaking these numbers down by discipline, we can see in Figure~\ref{fig:xlinglyhistbydisc} that there is no large difference, although mathematics papers tend to have a slightly higher portion of cross-lingual citations. The mean values for mathematics, physics and computer science are 0.090, 0.078, and 0.080 respectively.

Regarding prevalence we observe that in English papers in the disciplines of physics, mathematics, and computer science about 1 in 66 publications contains at least one explicitly marked citation to a non-English document. About two thirds of these citations are to Russian documents, although in the last years there is a downwards trend with regard to Russian and an upwards trend in citations to Chinese. Furthermore, cross-lingual citations appear about twice as often in mathematics compared to physics and computer science.
These observations suggest that while cross-lingual citations are not very frequent in general, they might be worth considering in applications dealing with specific disciplines and languages (e.g. citations to Russian in mathematics publications).

\subsection{Usage}

Regarding the usage of cross-lingual citations in English publications we analyze four different aspects. (1)~Whether or not self-citations are a driving factor, (2)~to what degree the geographical origin of a cross-lingual citation is correlated with the cited document's language, (3)~what function they serve, and (4)~what sentiment they express toward the cited document.

\subsubsection{Self-citation}\label{sec:selfcit}

To assess the relative degree of self-citation when referring to publications in other languages, we compare the ratio of self-citations in (a)~the \emph{cross-lingual citations} within the documents of the cross-lingual set, and (b)~the \emph{monolingual citations} within the documents of the cross-lingual set. Comparing two sets of citations from identical documents allows us to control for confounding effects such as author specific self-citation bias.

To determine self-citation, we rely on the author metadata in the MAG and therefore require both the citing and cited document of a reference to have a MAG ID. Within the cross-lingual set, this is the case for 3,370 cross-lingual references and 264,341 monolingual references. While at first, we strictly determine a self-citation by author IDs in the MAG being identical, manual inspection of matches and non-matches reveals, that author disambiguation within the MAG is somewhat lacking---that is, in a non-trivial amount of cases there are several IDs for a single author. We therefore measure self-citation by two metrics. A strict metric which only counts a match of MAG IDs, and a loose metric which counts an overlap of the sets of author names on both ends of the reference as a self-citation.

\begin{table}[tb]
\caption{Self-citations}
 \label{tab:selfcit}
  \centering
  \begin{small}
 \begin{threeparttable}
 \begin{tabular}{lrr}
 \toprule
   \ & \multicolumn{2}{c}{Self-citations} \\
   References to & loose & strict \\
   \midrule
   non-English & 19\% & 5\% \\
   English & 17.9\% & 11.3\% \\
   \bottomrule
 \end{tabular}
\end{threeparttable}
  \end{small}
\end{table}

Table~\ref{tab:selfcit} shows that going by the strict metric, self-citation is twice as common in monolingual citations. Applying the loose metric, however, self-citation appears to be slightly more common in cross-lingual citations. The larger discrepancy between the results of the strict and loose metric for cross-lingual citations suggests that authors publishing in multiple languages might be less well disambiguated in the MAG. With regard to self-citation being a motivating factor for cross-lingual citations---be it, for example, due to the need to reference one's own prior work---, we can note that our data does not suggest this to be the case. Authors using cross-lingual citations appear to be at least equally as likely to self-cite when referencing English works.

\subsubsection{Geographical Origin}

\begin{figure}[tb]
\centering
\includegraphics[width=\linewidth]{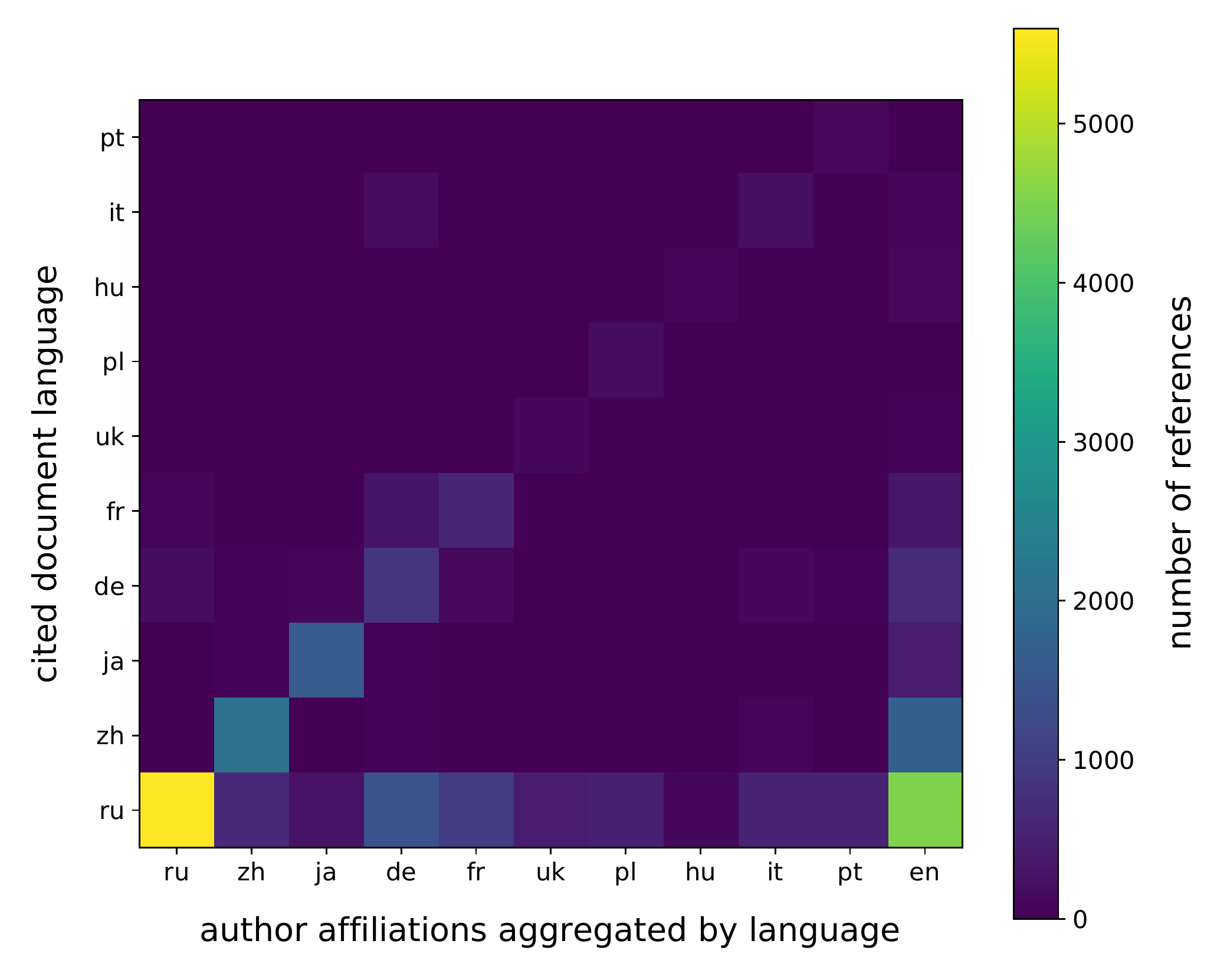}
\caption{Geographic origin of cross-lingual citations to the ten most cited languages (absolute count).} \label{fig:geo_abs}
%\end{figure}

%\begin{figure}[tb]
\centering
\includegraphics[width=\linewidth]{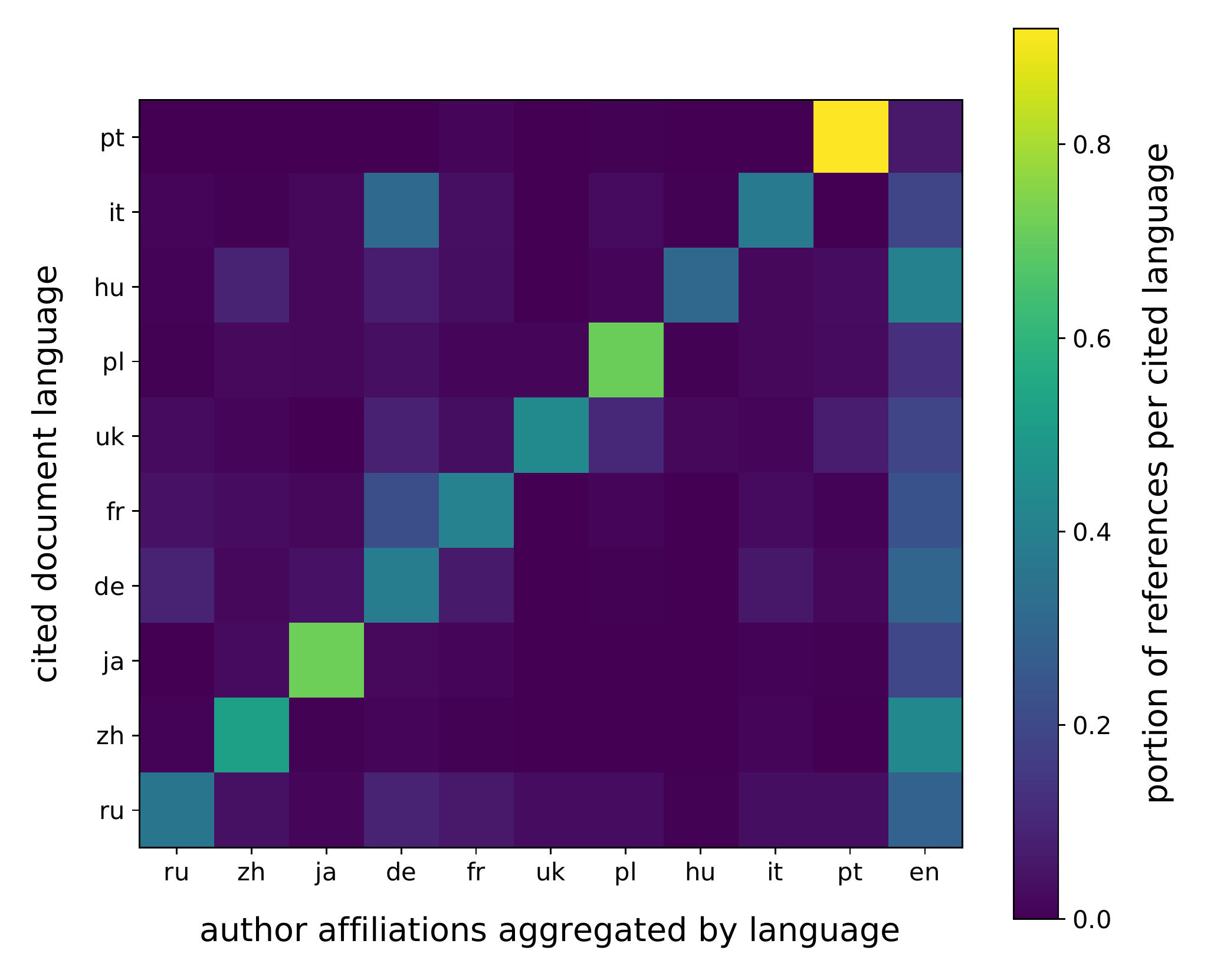}
\caption{Geographic origin of cross-lingual citations to the ten most cited languages(relative count).} \label{fig:geo_rel}
\end{figure}

In this section we analyze the geographical origin of cross-lingual citations. As a measure for geographical origin we use the country in which a citing author's affiliation is located. We refer to a citation as being to a ``local language'' or of ``local origin'', if the cited document's language is the most commonly spoken language in the affiliation's location. An example of this would be a researcher affiliated with a research institution located in Russia, being the author of a paper in which they cite a publication written in Russian.

For our analysis, we rely on author affiliation metadata in the MAG. We start off with all documents in the cross-lingual set that have a MAG ID.\footnote{I.e., documents for which we have MAG metadata (see Table~\ref{tab:dataused}).} From those, we select all which provide information on the authors' affiliations.\footnote{Because a single paper can have authors affiliated with institutions in different locations, we perform our analysis on a per author basis.} This leaves us with 7,522 out of 16,300 papers. To associate an author's affiliation with a language, we use the most commonly spoken language in the country or territory.\footnote{The association between affiliation and country is already given in the MAG. For data on language use per country we refer to the Unicode Common Locale Data Repository's territory-language information (see \url{https://unicode-org.github.io/cldr-staging/charts/latest/supplemental/territory_language_information.html}).} Grouping affiliations by language, we can then view the correlation of (a) cited languages and (b) language grouped affiliations in two ways. On the one hand, we can see for each cited language how many of the citations are of local origin---compared to, for example, from an English speaking country. On the other hand, we can see for each language group of affiliations how many cross-lingual citations are to a local language. Our results of this analysis are shown for the 10 most commonly cited languages in Figures~\ref{fig:geo_abs} and~\ref{fig:geo_rel}, and for all identified cited languages in Appendix~\ref{sec:apndx_geo}.

Figure~\ref{fig:geo_abs} shows citation numbers in absolute terms. Looking, for example, at citations to Russian publications (the bottom row of the figure), we can see that the largest amount of citations originates from Russian speaking countries (5,599 out of 18,672) followed by English speaking countries (4,535) and German speaking countries (1,427).

In Figure~\ref{fig:geo_rel} we show relative numbers per cited language. That is, the values of each \emph{row} add up to 1. Here we can see that citations to Japanese, Polish and particularly Portuguese appear to be of local origin comparatively often, with 68\% for Japanese, 64\% for Polish and 86\% for Portuguese. Overall we observe that cross-lingual citations are most often either of local origin or from an English speaking country. Evaluated over all languages, 37\% of cross-lingual citations are local (the diagonal in Figures~\ref{fig:geo_abs} and~\ref{fig:geo_rel}), while 26\% are from the Anglosphere (the ``en'' column in Figures~\ref{fig:geo_abs} and~\ref{fig:geo_rel}).

\begin{figure}[tb]
\centering
\includegraphics[width=\linewidth]{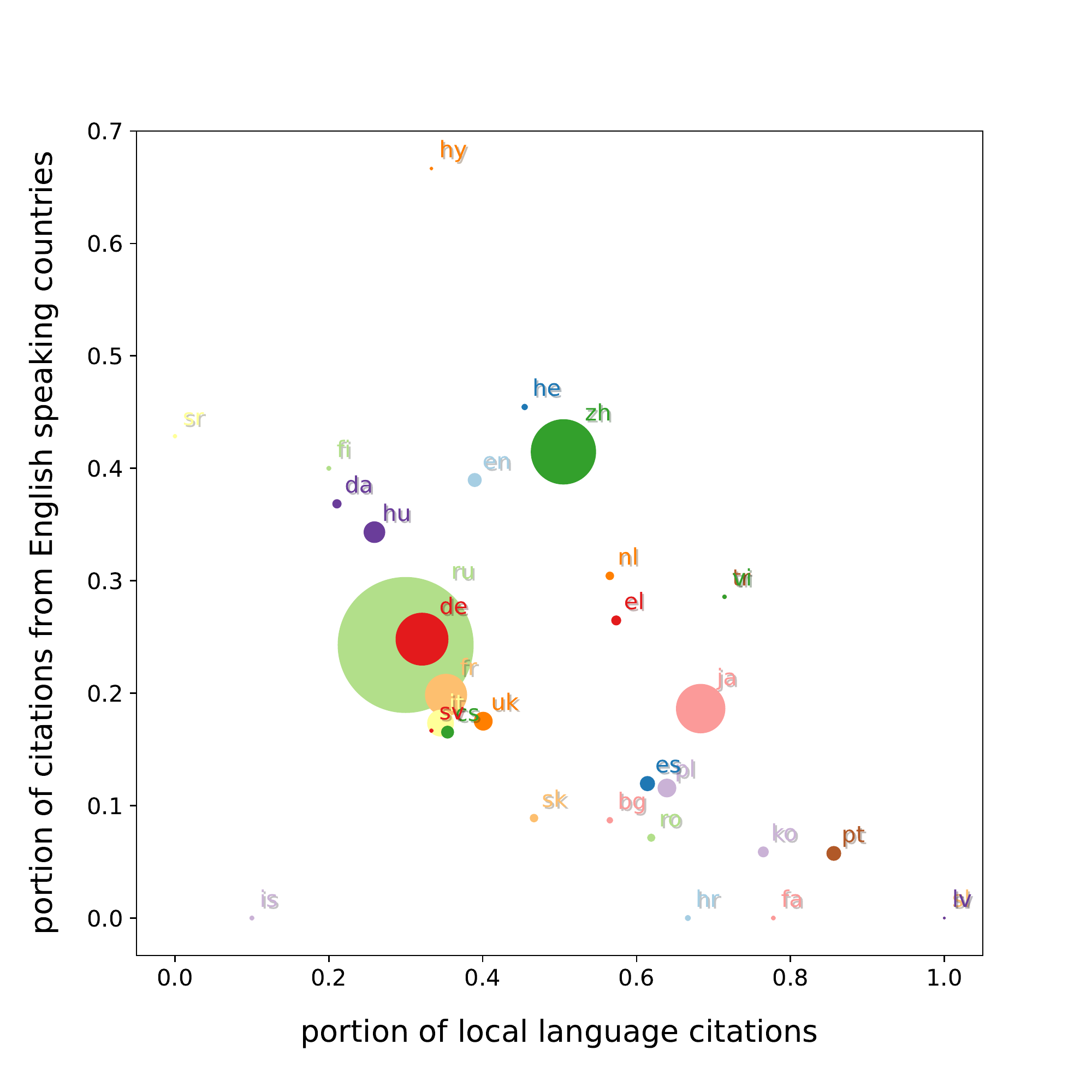}
\caption{Geographic origin of cross-lingual citations (local vs. English speaking countries). Marker size (surface area) indicates number of citations.} \label{fig:geo_localvsen}
\end{figure}

In Figure~\ref{fig:geo_localvsen} we jointly visualize how ``locally'' cited each language in our corpus is (x-axis) compared to which portion of citations originate from English speaking countries (y-axis). Overall, we observe larger variation on the ``locality'' dimension (values ranging from 0 to 1 with a variance of 0.058) than on the ``from English speaking countries'' dimension (values from 0 to 0.67 with a variance of 0.026). Looking at non-Latin script languages, we can see that Cyrillic script languages (e.g., Russian and Ukrainian) are less often of local origin than Asian languages (Chinese, Japanese, Korean) or languages written in Arabic script (Persian\footnote{While most varieties of Persian are written in a version of the Arabic script, there also exists varieties written in Cyrillic script~\cite{Megerdoomian2008}.}). Narrowing down on above-mentioned three Asian languages, we observe that for Chinese the relative portion of citations from English-speaking countries (0.41) is more than double of the same measure for Japanese (0.19), which is more than triple the value for Korean (0.06). The comparatively high ratio for Chinese (not just among Asian languages but overall\footnote{The overall comparison has, however, to be done keeping the limitations described in Section~\ref{sec:ident} in mind.}) could be taken as an indication for two phenomena: first, an increased relevance of publications written in Chinese (i.e., a higher necessity to cite) and second, an increased rate of scholars able to read Chinese in English speaking country research institutions (i.e., a higher probability of the ability to cite).

\subsubsection{Citation Intent and Sentiment}

To assess whether or not cross-lingual citations tend to serve a different purpose than their monolingual counterpart, and whether or not authors have a different disposition toward cited works, we analyze the in-text citations (see Figure~\ref{fig:terminology}) in our corpus.

The analysis of in-text citations---commonly referred to as citation context analysis---is concerned with the textual context of citations~\cite{Hernandez2016}. Two tasks in citation context analysis are the classification of citation intent (also referred to as citation function) and citation sentiment (also referred to as citation polarity)~\cite{Hernandez2016}. Citation intent can reveal why an author added a reference, while the citation sentiment can give insight into the author's disposition toward that reference. Both citation intent and sentiment have been used in a number of diverse tasks, such as classification \cite{Jurgens2018,Cohan2019,Beltagy2019}, summarization \cite{Cohan2015}, and citation recommendation \cite{Faerber202x}. For citation intent, many schemes have been proposed to classify different functions, ranging from fine-grained to coarse-grained schemes. A partial overview of these can be found in Hernández-Alvarez~\cite{Hernandez2016}, Jurgens et al.~\cite{Jurgens2018}, Cohan et al.~\cite{Cohan2019}, and Lauscher et al.~\cite{multicite-lauscher-2021}. These schemes, however, are often domain-specific and too fine-grained~\cite{Cohan2019}. Jurgens et al.~\cite{Jurgens2018} proposed a unified scheme of previous work (with six categories), while Cohan et al.~\cite{Cohan2019} proposed a more generalized scheme (with three categories) that works for multiple domains. Recently, Lauscher et al. ~\cite{multicite-lauscher-2021} expanded these schemes to multi-sentence and multi-label citation contexts. Given the number of diverse domains on arXive, we adopt the general scheme by Cohan et al.~\cite{Cohan2019}. For citation sentiment, a three category scheme (\textit{positive}, \textit{negative}, or \textit{neutral}) is widely adopted~\cite{Athar2011,Abujbara2013,Hernandez2016}. Previous approaches to citation intent and sentiment classification have used either hand-crafted rules or classical machine learning models~\cite{Abujbara2013,Jurgens2018}, while more recent approaches using deep learning and word embeddings have demonstrated significant improvements in performance~\cite{Cohan2019,Beltagy2019,multicite-lauscher-2021}.

For our analysis, we create two, equally-sized sets of in-text citations. The \textit{in-text x-ling} set (cross-lingual) and the \textit{in-text mono} set (monolingual). In the following we describe the creation of both sets, the classifier model training, and our results for citation intent and sentiment classification.

\begin{figure}[tb]
\centering
\includegraphics[width=0.8\linewidth]{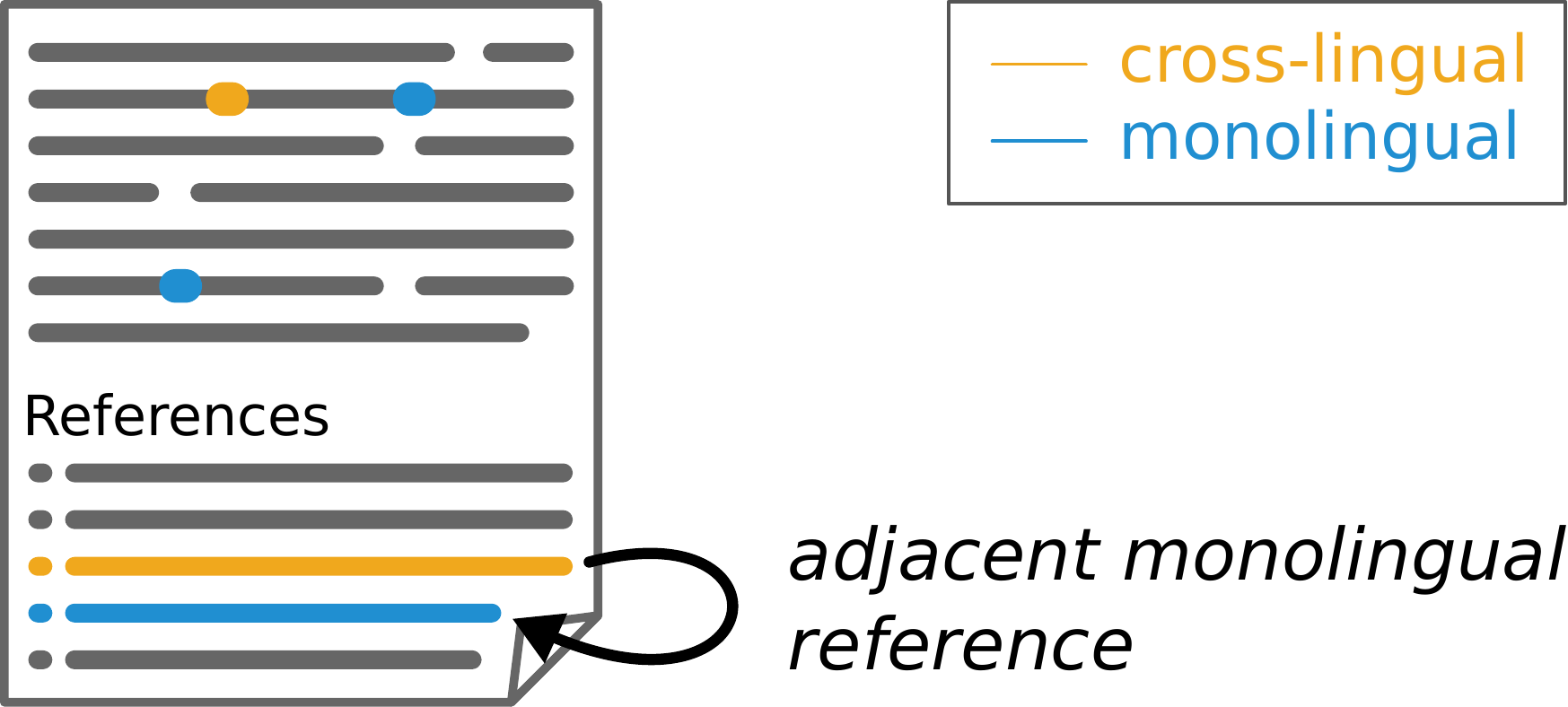}
\caption{Schematic explanation of an adjacent monolingual reference.} \label{fig:adjrefsampling}
\end{figure}

\textsc{Data Preparation}
For the \textit{in-text x-ling} set we determine all in-text citations associated with the references in the cross-lingual set. This yields 45,516 in-text citations for our 33,290 cross-lingual references. The \textit{in-text mono} set is then created by extracting in-text citations associated with adjacent monolingual references. We illustrate this process in Figure~\ref{fig:adjrefsampling}, showing a paper with a single cross-lingual reference for which, accordingly, a single adjacent monolingual reference would be determined and its associated in-text citations (indicated by the two blue markers above) extracted. For \textit{in-text mono} we extract 53,177 in-text citations (i.e., on average more in-text citations per reference) which we reduce to 45,516 through stratified sampling. By sourcing our monolingual in-text citations for comparison from the same papers, we avoid confounding effects such as author specific differences in citation styles.

As a citing sentence can contain more than one citation marker, it is possible that the in-text citations associated with two adjacent reference section entries appear within the same sentence (e.g., as indicated in the second ``text'' line in Figure~\ref{fig:adjrefsampling}). This is the case for 10,454 of the in-text citations we extracted (i.e., these appear in both sets). We define them as a third set called \textit{mixed}, leaving \textit{in-text x-ling} and \textit{in-text mono} at 35,062 items each.

\begin{table}
\caption{Class distribution and evaluation details for the model training.}
 \label{tab:classdist}
  \centering
  \begin{small}
 \begin{threeparttable}
 \begin{tabular}{lrrrrrr}
 \toprule
   Data set & Class & Inst.\tnote{a}\hphantom{\,} & P\tnote{b}\hphantom{\,} & R\tnote{c}\hphantom{\,} & F1\tnote{d} \\
   \midrule
   \multirow{3}{*}{SciCite} & Backgr. & 6,375 (58) & 86\% & 93\% & \multirow{3}{*}{86.6\%}\\
                            & Method & 3,154 (29) & 91\% & 82\%\\
                            & Result & 1,491 (13) & 86\% & 83\% \\ \midrule\midrule
  \multirow{3}{*}{Athar} & Neutral & 6,901 (87) & 91\% & 98\% & \multirow{3}{*}{67.9\%} \\
                          & Positive & 761 (10) & \textbf{80\%} & 42\% \\
                          & Negative & 265 (3) & 50\% & 29\% \\ \midrule
  \multirow{3}{*}{Athar$^{\dagger}$} & Neutral & 265 (33) & 77\% & 59\% & \multirow{3}{*}{67.7\%} \\
                          & Positive & 265 (33) & 59\% & 59\% \\
                          & Negative & 265 (33) & \textbf{65\%} & \textbf{94\%} \\ \midrule
    \multirow{2}{*}{Athar$^{\mathsection}$} & Neutral & 6,901 (90) & \textbf{96\%} & \textbf{97\%} & \multirow{2}{*}{82.5\%} \\
                            & Positive & 761 (10) & 69\% & 68\% \\ \midrule
    \multirow{2}{*}{Athar$^{\ddagger}$} & Neutral & 761 (50) & 85\% & 69\% & \multirow{2}{*}{80.2\%} \\
                            & Positive & 761 (50) & 78\% & \textbf{90\%} \\
   \bottomrule
 \end{tabular}
 \begin{tablenotes}
    \item[a] Inst. = Number of instances for training and evaluation
    \item[\hphantom{a}] \ \hphantom{Inst. =}(percentage in brackets)
    \item[b] P = Precision score on test set
    \item[c] R = Recall score on test set
    \item[d] F1 = F1-macro score on test set
    \item $^{\dagger}$ = Under-sampled
    \item $^{\mathsection}$ = No \textit{Negative} class
    \item $^{\ddagger}$ = Under-sampled \& no \textit{Negative} class
 \end{tablenotes}
\end{threeparttable}
  \end{small}
\end{table}

\textsc{Model Training}
Training data for citation sentiment and intent classification regarding papers cannot easily be crowdsourced, because domain knowledge is needed for annotation. As a consequence, available data sets are comparatively small. We identify SciCite~\cite{Cohan2019} for citation intent and the data set proposed by Athar~\cite{Athar2011} for citation sentiment as most appropriate for our purposes.

\begin{itemize}
\item SciCite contains 11,020 citations that originate from the Semantic Scholar corpus, which covers several disciplines such as computer science, molecular biology, microbiology and neuroscience~\cite{Ammar2018}. Citations in SciCite are labeled regarding their intent across three categories, namely \textit{Background}, \textit{Method}, and \textit{Result}. The class distribution can be seen in Table \ref{tab:classdist}. We select the data set because it is currently the largest available, and classifiers trained on the data set achieve good performance.

\item The data set created by Athar contains 8,736 annotated citations from 310 research papers. To the best of our knowledge, it is the largest citation sentiment data set currently available. Following~\cite{Mercier2021}, we manually remove 809 items from the data set that are either duplicates or too short to be accurately evaluated regarding their sentiment. The resulting data set, which we refer to as \textit{Athar} from hereon, contains 7,927 citations annotated with one of the three labels \textit{Negative}, \textit{Neutral}, and \textit{Positive}. Citations labeled \textit{Negative} and \textit{Positive} are comparably infrequent in the corpus (see Table \ref{tab:classdist}), which makes classifying them more difficult. As possible mitigation strategies, we consider the following options.
    \begin{itemize}
    \item Athar$^{\dagger}$: balancing the data by under-sampling.
    \item Athar$^{\mathsection}$: removing the \textit{Negative} class, as its low performance (see Table~\ref{tab:classdist}) puts its informativeness into question.
    \item Athar$^{\ddagger}$: both of the aforementioned.
    \end{itemize}
\end{itemize}

For each of our classification models, we fine-tune SciBERT~\cite{Beltagy2019}, a pre-trained language model for scientific text that achieves state-of-the-art performance on sentence classification tasks.

Our evaluation results are shown in Table~\ref{tab:classdist}. On both SciCite and Athar our models perform on par with the best performing models presented in their respective publications. For citation intent, we achieve an F1 score of 86.6\% and relatively similar performance across classes. For citation sentiment, we achieve an F1 score of 67.9\% on the original Athar data set. Two of our three class imbalance mitigation strategies (Athar$^{\mathsection}$ and Athar$^{\ddagger}$) result in an increase in the F1 score to over 80\%. Of those two we decide to use the model trained on Athar$^{\ddagger}$. While training on Athar$^{\mathsection}$ gives us a slightly higher F1 score, the model trained on Athar$^{\ddagger}$ achieves high precision and recall for positive citations---which are presumably less common---while also maintaining good performance for neural citations.
Implementation details for the model training can be found in Ap\-pen\-dix~\ref{app:classifcation}.

\begin{table}[tb]
\caption{Citation intent and sentiment classification results for cross-lingual, monolingual, and mixed in-text citations. (Values are the number of citations per class followed by the percentage in brackets.)}
 \label{tab:citationclassificationdist}
  \centering
  \begin{small}
 \begin{threeparttable}
 \begin{tabular}{lrrr}
 \toprule
  Data set\hphantom{\ } & Background\hphantom{\ } & Method\hphantom{\ } & Result\hphantom{\ }\\
  \midrule
    \textit{x-ling} & 26,443 (75.4) & 7,749 (22.1) & 870 (2.5) \\
    \textit{mono} & 26,232 (74.8) & 7,801 (22.2) & 1,029 (2.9) \\
    %\midrule
    \textit{mixed} & 7,688 (73.5) & 2,503 (23.9) & 263 (2.5) \\
  \midrule
  \midrule
  & Neutral\hphantom{\ } & Positive\hphantom{\ } & Negative\hphantom{\ }\\
  \midrule
    \textit{x-ling*} & 34,100 (97.3) & 787 (2.2) & 175 (0.5) \\
    \textit{mono*} & 33,792 (96.4) & 1,037 (3.0) & 233 (0.7) \\
    \textit{mixed*} & 10,049 (96.1) & 362 (3.5) & 43 (0.4) \\ \midrule
    \textit{x-ling$^{\ddagger}$} & 22,275 (63.5) & 12,787 (36.5) & \\
    \textit{mono$^{\ddagger}$} & 21,825 (62.3) & 13,237 (37.8) & \\
    \textit{mixed$^{\ddagger}$} & 6,547 (62.6) & 3,907 (37.4) & \\
  \bottomrule
 \end{tabular}
  \begin{tablenotes}
    \item * = Classified using the model trained on Athar
    \item $^{\ddagger}$ = Classified using the model trained on Athar$^{\ddagger}$
    % \item*** = Unbalanced 
 \end{tablenotes}
\end{threeparttable}
  \end{small}
\end{table}

\begin{figure}[tb]
\centering
\includegraphics[width=\linewidth]{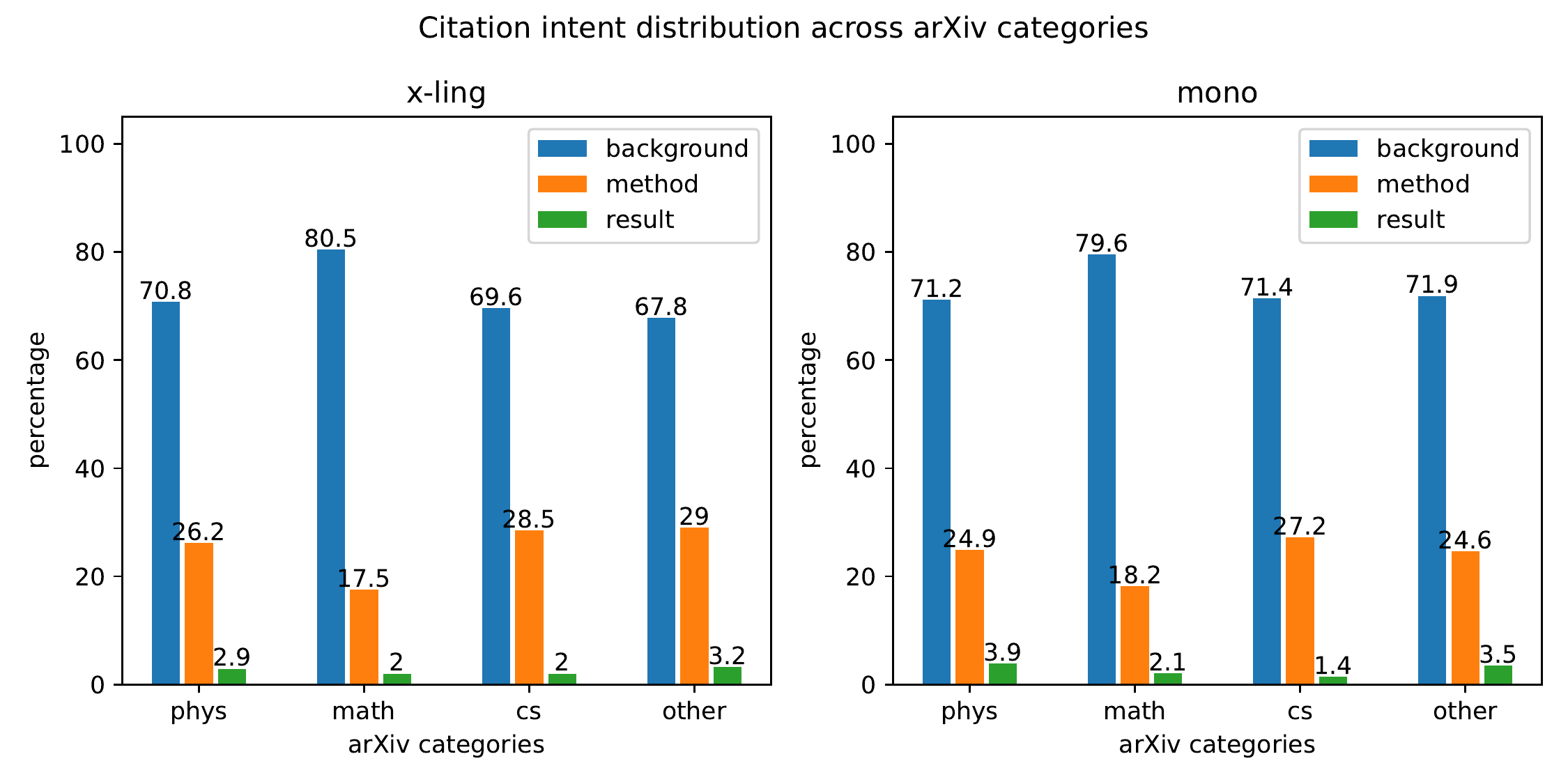}
\caption{Comparison of citation intent distribution across arXiv categories for \textit{in-text x-ling} (left) and \textit{in-text mono} (right).} \label{fig:intentarxivcategories}
\end{figure}

\textsc{Classification Results}
Based on above evaluation we proceed by using our models trained on SciCite, Athar, and Athar$^{\ddagger}$ to classify the intent and sentiment of citations in \textit{in-text x-ling} and \textit{in-text mono}.
In Table~\ref{tab:citationclassificationdist}, we show the classification results for citation intent (top half) and sentiment (bottom half). The classifiers trained on SciCite and Athar appear to amplify the unbalanced data distribution they were trained on to some degree. Comparing the sentiment classifiers trained on the original Athar and balanced Athar$^{\ddagger}$ data set, we see that citations classified as \textit{Positive} increase from around 3\% to almost 38\%. We take this as a clear sign that reliably distinguishing neutral from positive citations remains a challenge even with state-of-the art models and training data.

Comparing our results across the data sets \textit{in-text x-ling}, \textit{in-text mono}, and \textit{in-text mixed} we see that in terms of both intent and sentiment class distributions are similar. Taking a closer look at citation intent across the scientific disciplines,\footnote{We do not evaluate citation sentiment here due to the lacking performance of the sentiment classifiers.} we can see in Figure~\ref{fig:intentarxivcategories} that the distributions are overall comparable among disciplines and between cross- and monolingual citations, with mathematics showing a slightly higher use of background citations.

Overall, our results for citation sentiment and intent show no distinct differences between cross- and monolingual citations. This can be taken as an indication for two things. First, that authors cite existing literature with a certain intent and sentiment \emph{regardless} of the cited work's language. Second, that cross-lingual---while occurring less frequent---serve the same functions as monolingual citations and are therefore not less significant.

\subsection{Impact}
Regarding the impact of cross-lingual citations we analyze whether cross-lingual citations in English papers are seen as an ``acceptable'' practice, whether or not they pose a particular challenge for citation data mining, and their potential impact on the success of the paper they're part of. Our results concerning these three aspects are described in the following sections.

\subsubsection{Acceptance}

To assess the acceptance of cross-lingual citations by the scientific community---that is, whether or not non-English publications are deemed ``citable''~\cite{Lillis2010}---we analyze papers in our data that have both a preprint version as well as a published version (in a journal or conference proceedings) dated later than the preprint. This is the case for 2,982 papers. For each preprint-published paper pair, we check if there is a difference in cross-lingual citations. This gives an indication of how the process of peer review affects cross-lingual citations.
We perform a manual as well as an automated analysis.\footnote{Full evaluation details can be found at \url{https://github.com/IllDepence/cross-lingual-citations-from-en}.}

For the manual evaluation, we take a random sample of 100 paper pairs. We then retrieve a PDF file of both the preprint and the published version, and manually compare their reference sections. For the automated evaluation, we find that 599 of the 2.9k paper pairs have PDF source URLs given in the MAG. After automatically downloading these and parsing them with GROBID, we are left with 498 valid sets of references. For these, we identify explicitly marked cross-lingual references as described in Section~\ref{sec:datacollection} and calculate their differences.

Table~\ref{tab:peerreview} shows the results of our evaluations. In both, cross-lingual citations are more often removed than added, but in the majority of cases left intact. The larger volatility in the automated evaluation is likely due to parsing inconsistencies of GROBID. Our findings complement those of Lillis et al.~\cite{Lillis2010}, who, analyzing psychology journals, observe \textit{``some evidence that gatekeepers [...] are explicitly challenging citations in other languages.''} For the fields of physics, mathematics, and computer science, we find no clear indication of a consistent in- or decreasing effect of the peer review process on cross-lingual citations.

\begin{table}
\caption{Changes in cross-ling. cit. between preprints and published papers}
 \label{tab:peerreview}
  \centering
  \begin{small}
 \begin{threeparttable}
 \begin{tabular}{lrrrrrr}
 \toprule
   Evaluation & \#Pairs & \#Inc.\tnote{a} & \#Dec.\tnote{b} & Mean\tnote{c} & SD\tnote{c} \\
   \midrule
   Manual & 100 & 4 & 7 & -0.02 & 0.529 \\
   Automated & 498 & 33 & 70 & -0.12 & 0.821 \\
   \bottomrule
 \end{tabular}
 \begin{tablenotes}
    \item[a] Inc. = Increased
    \item[b] Dec. = Decreased
    \item[c] of the differences in the amount of cross-lingual citations
 \end{tablenotes}
\end{threeparttable}
  \end{small}
\end{table}

\subsubsection{Impact on Paper Success}

To get an indication of whether or not an English paper's success is influenced by the fact that it contains citations to non-English documents, we compare our cross-lingual set with the random set (cf. Table~\ref{tab:datasets}). For both sets we first determine the number of papers that in the MAG metadata have a published version (journal or conference proceedings) in addition to the preprint on arxiv.org. That is, we assume that papers which only have a preprint version did not make it through the peer review process. Using this measure, we observe 9,390 of 16,224 (57.88\%) successful papers in the cross-lingual set, and 10,966 of 16,378 (66.96\%) successful papers in the random set. Unsurprisingly, due to the higher ratio of published versions, the papers in the random set are also cited more. Table~\ref{tab:citcounts} shows a comparison of the average number of citations that documents in both sets received. Due to the high standard deviation in the complete sets, we also look at papers which received between 1 and 100 citations, which are comparably frequent in both sets. As we can see, in the unfiltered as well as the filtered case, documents with cross-lingual citations tend to be cited a little less. Because here we can only control for the distribution of papers across years and disciplines, and not for individual authors (as we did in the Section~\ref{sec:selfcit}), there might be various confounding factors involved.

\begin{table}[tb]
\caption{Comparison of citations received}
 \label{tab:citcounts}
  \centering
  \begin{small}
 \begin{threeparttable}
 \begin{tabular}{llrr}
 \toprule
   \multicolumn{2}{l}{Filter criterion} & Cross-lingual set & Random set \\
   \midrule
   - & \#Docs & 16,300 & 16,464 \\
   \ & Mean \#cit & 13.7 & 18.2 \\
   \ & SD & 75.0 & 51.7 \\
   \midrule
   $1\le \#cit$ & \#Docs & 12,074 & 12,852 \\
   and & Mean \#cit & 12.0 & 15.1 \\
   $\#cit\le 100$ & SD & 15.8 & 18.4 \\
   \bottomrule
 \end{tabular}
\end{threeparttable}
  \end{small}
\end{table}

\subsubsection{Impact on Citation Data Mining}

To assess if cross-lingual citations pose a particular challenge for scholarly data mining---and are therefore likely to be underrepresented in scholarly data---, we compare the ratio of references that could be resolved to MAG metadata records for the cross-lingual set and the whole unarXive data set. Of the 39M references in unarXive 42.6\% are resolved to a MAG ID. For the complete reference sections of the papers in the cross-lingual set (i.e., references to both non-English and English documents) the number is 45.7\% (290,421 of 635,154 references). Looking only at the cross-lingual citations, the success rate of reference resolution drops to 11.2\% (3,734 of 33,290 references). We interpret this as a clear indication that resolving cross-lingual references is a challenge. Possible reasons for this are, for example:
\begin{enumerate}
\item A lack of language coverage in the target data set.\\For example, if the target data set only contains records of English papers, references to non-English publications cannot be found within and resolved to that target data set.
\item Missing metadata in the target data set.\\For example, when there is a primary non-English as well as an alternative English title of a publication, only the former is in the target data set's metadata, but the latter is used in the cross-lingual reference.
\item The use of a title translated ``on the fly.''\\If a non-English publication has no alternative English title, a self translated title in a reference cannot be found in any metadata. To give an example, reference 14 in \texttt{arXiv:1309.1264} titled \textit{``Hierarchy of reversible logic elements with memory''} is only found in metadata\footnote{See \url{http://hdl.handle.net/2433/172983}.} as \ja{記憶付き可逆論理素子の能力の階層構造について}.
\item The use of a title transliterated ``on the fly.''\\Similar to an unofficial translated title, if a title is transliterated and this transliteration is not existent in metadata, the provided title is not resolvable. A concrete example of this is the third reference in \texttt{arXiv:cs/9912004} titled \textit{``Daimeishi-ga Sasumono Sono Sashi-kata''} which is only found in metadata\footnote{See \url{https://ci.nii.ac.jp/naid/10008827159/}.} as \ja{代名詞が指すもの,その指し方}.
\end{enumerate}

Cases 4 and especially 3 additionally impose a challenge on human readers, as the referred documents can only be found by trying to translate or transliterate back to the original. References to non-English documents which do not have an alternative English title should therefore ideally include enough information to (a) identify the referenced document (i.e., at least the original title), and (b) a way for readers not familiar with the cited document's language to get an idea of what is being cited (e.g., by adding a freely translated English title).\footnote{And example for this can be found in reference 15 in \texttt{arXiv:1503.05573}: ``\foreignlanguage{russian}{Шафаревич И. Р. Основы алгебраической геометрии// МЦНМО, Москва, 2007.} (English translation: Shafarevich I.R. Foundations of Algebraic Geometry// MCCME, Moscow. 2007).''} There are, however, situations where an original title cannot be used. Documents in PubMed Central, for example, cannot contain non-Latin scripts,\footnote{See \url{https://www.ncbi.nlm.nih.gov/pmc/about/faq/\#q16}.} meaning that references to documents in Russian, Chinese, Japanese, etc. which do not have alternative English titles are inevitably a challenge for both human readers as well as data mining approaches, unless there is a DOI, URL, or similar identifier that can be referred to.

In light of this, taking a closer look at the 88.8\% of unmatched references in the cross-lingual set broken down by languages, we note the following matching failure rates for the five most prevalent languages: Russian: 88.6\%, Chinese: 87.0\%, Japanese: 91.0\%, German: 85.4\%, and French: 83.2\%. While all of these are high, the numbers for the three non-Latin script languages are noticeably higher than those of German and French. As can be seen with the task of resolving references---and as also indicated through our self-citation data shown in Table~\ref{tab:selfcit}---cross-lingual citations do pose a particular challenge for scholarly data mining.

\section{Discussion and Conclusion}
\label{sec:conclusion}

Utilizing two large data sets, unarXive and the MAG, we performed a large-scale analysis of citations from English papers to non-English language publications (i.e., cross-lingual citations). The data analyzed spans over one million citing publications, 3 disciplines, and 27 years. We gained insights into cross-lingual citations' prevalence, usage and impact.

Recapitulating our key results, we find that citations to non-Latin script languages can reliably be identified by a \textit{``(in $<$Language$>$)''} marker, which enables automated identification in large corpora. Between the disciplines of physics, mathematics, and computer science, cross-lingual citations appear twice as often in mathematics papers compared to the remaining two fields. Over the course of time we see a downwards trend in citations to Russian and an upwards trend for citations to Chinese. In general, cross-lingual citations are more often of linguistically local origin than originating from English speaking countries. Citations to Chinese, however, are about twice as likely to come from the Anglosphere than citations to other languages. Concerning authors citing behavior, we observe no remarkable differences between cross- and monolingual citations in terms of self-citations, intent, and sentiment. We also see no clear indication for gatekeeping of cross-lingual citations through the process of peer review. As for the impact of cross-lingual citations on a paper's success, we only get inconclusive results. Finally, we see clear indicators that cross-lingual citations pose challenges for scholarly data mining, such as a lower likelihood to resolve a cited document due to more complex metadata (e.g., publications having two titles, a primary non-English and an alternative English title) and shortcomings in data integration (e.g., with local citation indices).

Through our preliminary analyses (see Section~\ref{sec:ident}) we identify challenges in reliably assessing cross-lingual citations to Latin script languages, preventing automated identification in large corpora. These insights can facilitate future efforts in overcoming the identified challenges. Our detailed findings regarding prevalence can help identify scenarios, in which a dedicated effort to take into account cross-lingual citations is warranted. For example, a citation driven analysis of research trends in mathematics might benefit from being able to track ``citation trails'' into the realm of Russian publications. Lastly, due to the large scale of our investigation, the use of our collected data for machine learning based applications such as cross-lingual citation recommendation is possible.

As our analysis is based on explicit language markers of cited documents, which has shown to be reliable for non-Latin script languages but only capture a small fraction of citations to Latin script languages, we want to investigate further methods for identifying cross-lingual citations, to be able perform more exhaustive analyses. Furthermore, our corpus covers publications from the fields of physics, mathematics, and computer science. While arxiv.org has extensive coverage of physics and mathematics, the share of computer science publications is currently still in a phase of rapid growth. We therefore want to expand our investigation regarding computer science publications to get more representative results, but also include additional disciplines not covered so far.
Lastly, we would like to conduct complementary analyses of cross-lingual citations from non-English to English. These might be more challenging to perform on a large scale, because non-English scholarly data is not as readily available. However, such analyses are also likely to yield insights with a larger impact, as citing English language publications is rather common in other languages.

\section*{Author Contributions}  % cf. https://casrai.org/credit/
Tarek Saier: Conceptualization, Data curation, Formal analysis, Investigation, Methodology, Software, Visualization, Writing -- original draft (lead), Writing -- review \& editing. Michael F{\"a}rber: Supervision, Writing -- review \& editing. Tornike Tsereteli: Formal analysis, Software, Writing -- original draft (support).

\begin{acknowledgements}
We thank Irma Suppes for supporting manual data labeling and language identification tasks.
\end{acknowledgements}

\begin{appendices}

\section{Geographic origin of all cited non-English languages}\label{sec:apndx_geo}

\begin{figure*}[tb]
\centering
\includegraphics[width=\textwidth]{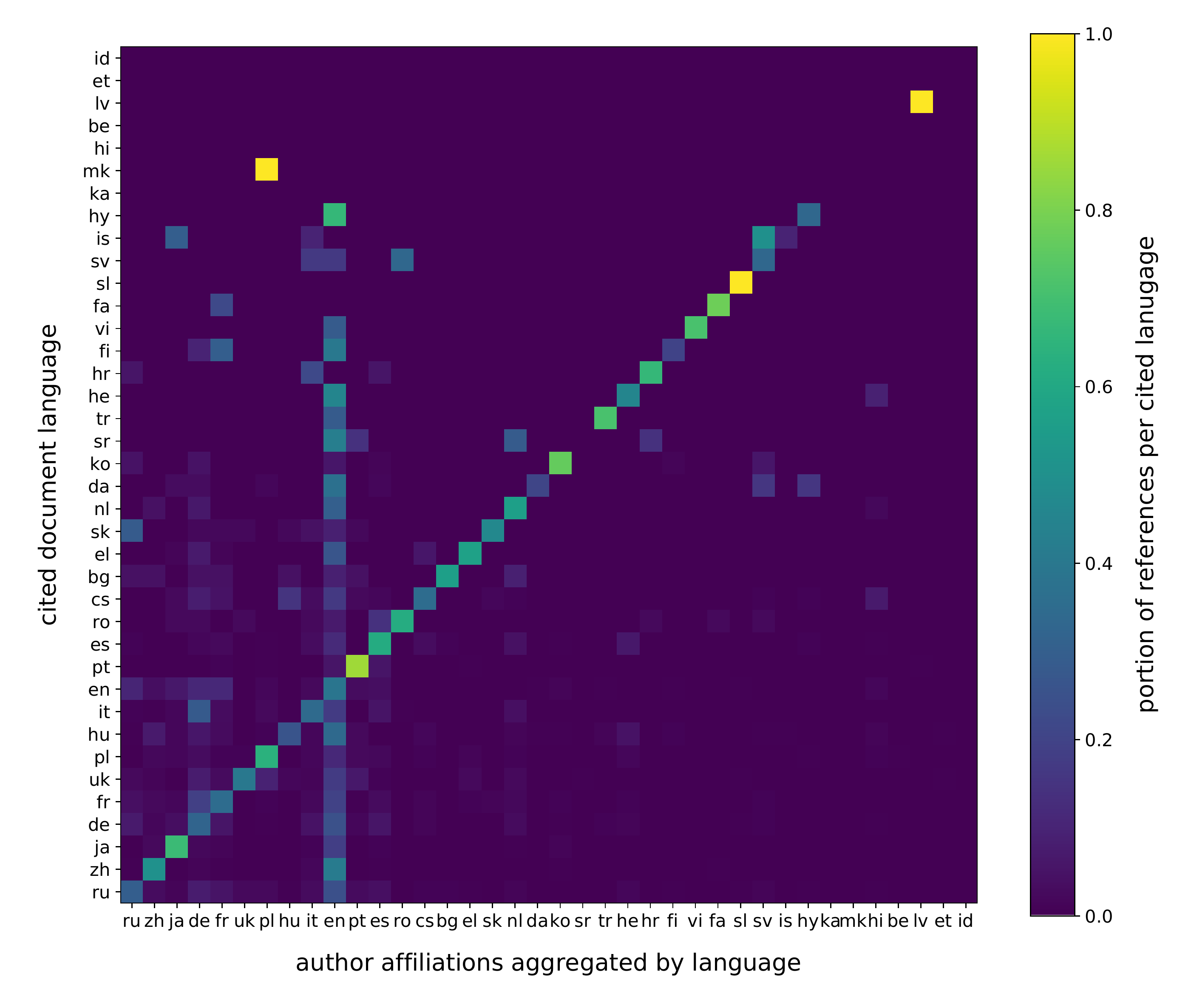}
\caption{Geographic origin of cross-lingual citations (relative count).} \label{fig:geo_full}
\end{figure*}

In Figure~\ref{fig:geo_full} we show the geographic origin of cross-lingual citations in relative terms per cited language (i.e., the numbers of each \emph{row} add up to 1). The distinct diagonal of the matrix and the horizontal line for affiliations in English speaking countries reflect the fact that most cross-lingual citations are either to a local language or originate from an English speaking country. Among cited languages with a low number of total occurrences we can furthermore see a few cases showing unusual distributions, such as a single citation to Macedonian from an author affiliated with a Polish institution, or citations to Icelandic, where a single one originates from Iceland, while the remaining nine originate from institutions in countries where Japanese (3), Italian (1), and Swedish (5) are the most common language.

\section{Citation Intent and Sentiment Classification}
\label{app:classifcation}

For the model training of both citation intent classification and citation sentiment classification, we fine-tune SciBERT uncased\footnote{See \url{https://huggingface.co/allenai/scibert_scivocab_uncased}.} using the following model configuration shown in Table~\ref{tab:modelconf}.

\begin{table}
\caption{Model configuration used for training}
 \label{tab:modelconf}
  \centering
  \begin{small}
 \begin{threeparttable}
 \begin{tabular}{lr}
 \toprule
   Hyperparameter & value \\
   \midrule
  attention\_probs\_dropout\_prob &  0.1 \\
  gradient\_checkpointing &  false \\
  hidden\_act &  gelu \\
  hidden\_dropout\_prob &  0.1 \\
  hidden\_size &  768 \\
  initializer\_range &  0.02 \\
  intermediate\_size &  3072 \\
  layer\_norm\_eps &  1e-12 \\
  max\_position\_embeddings &  512 \\
  model\_type &  bert \\
  num\_attention\_heads &  12 \\
  num\_hidden\_layers &  12 \\
  pad\_token\_id &  0 \\
  position\_embedding\_type &  absolute \\
  transformers\_version &  4.4.2 \\
  type\_vocab\_size &  2 \\
  use\_cache &  true \\
  vocab\_size &  31090 \\
   \bottomrule
 \end{tabular}
\end{threeparttable}
  \end{small}
\end{table}

For determining the citation intent, we use the train, validation, and test split provided by the SciCite data set\footnote{See \url{https://huggingface.co/datasets/scicite}.} (train: 74\%, val: 8.3\%, test: 16.9\%). For citation sentiment, we split the Athar data set into train, validation, and test sets into 80\%, 10\%, and 10\%, respectively.

\end{appendices}

\bibliographystyle{spmpsci}
\bibliography{xling_en}

\end{document}